\documentclass[runningheads]{llncs}
\usepackage[numbers,sort&compress]{natbib}
\usepackage{hyperref}
\usepackage{url}
\usepackage{graphicx}
\usepackage{wrapfig}
\usepackage{xcolor}
\usepackage[frozencache]{minted}
\setminted{
    linenos,
    frame=lines,
    fontfamily=tt
}
\usepackage[T1]{fontenc}
\usepackage{lmodern}
\usepackage{amsmath,amssymb,amsfonts}
\usepackage{algorithm}
\usepackage{algpseudocode}

\newcommand{\eg}{e.\,g.,\ }
\newcommand{\ie}{i.\,e.,\ }
\DeclareMathOperator{\BIN}{\mathtt{BIN}}
\DeclareMathOperator{\MBS}{\mathtt{MBS}}
\DeclareMathOperator{\XOR}{\oplus}

\begin{document}
\title{Accelerating quantum many-body configuration interaction with directives}
\author{Brandon Cook\inst{1} \and
  Patrick J. Fasano\inst{2}\orcidID{0000-0003-2457-4976} \and
  Pieter Maris\inst{3} \and
  Chao Yang\inst{4} \and
  Dossay Oryspayev\inst{5}}
\authorrunning{B. Cook et al.}
\institute{National Energy Research Scientific Computing Center, Lawrence Berkeley National Laboratory, Berkeley, CA, USA %
  \email{bgcook@lbl.gov}\\
  \and
  Department of Physics, University of Notre Dame, Notre Dame, IN, USA
  \email{pfasano@nd.edu}\\
  \and
  Department of Physics and Astronomy, Iowa State University, Ames, IA, USA
  \email{pmaris@iastate.edu}\\
  \and
  Computational Research Division, Lawrence Berkeley National Laboratory, Berkeley, CA, USA
  \email{cyang@lbl.gov}\\
  \and
  Computational Science Initiative, Brookhaven National Laboratory, Upton, NY, USA
  \email{doryspaye@bnl.gov}\\
}
\maketitle              %
\begin{abstract}
  Many-Fermion Dynamics---nuclear, or MFDn, is a configuration interaction (CI)
  code for nuclear structure calculations. It is a platform-independent
  Fortran~90 code using a hybrid MPI+X programming model. For CPU platforms the
  application has a robust and optimized OpenMP implementation for shared memory
  parallelism. As part of the NESAP application readiness program for NERSC's
  latest Perlmutter system, MFDn has been updated to take advantage of
  accelerators. The current mainline GPU port is based on OpenACC. In this work
  we describe some of the key challenges of creating an efficient GPU
  implementation. Additionally, we compare the support of OpenMP and OpenACC on
  AMD and NVIDIA GPUs.
\end{abstract}

\section{Introduction}

Many-Fermion Dynamics---nuclear, or MFDn, is a configuration interaction (CI)
code for \textit{ab initio} nuclear physics calculations.  It calculates the
approximate many-body wave function of self-bound atomic nuclei, starting from
two- and three-nucleon interactions.  In these calculations, the nuclear
many-body Hamiltonian is represented as a large sparse symmetric matrix in
configuration space.  The lowest eigenvalues of this matrix correspond to the
energy levels of the low-lying spectrum, and the eigenvectors represent the
corresponding many-body wave function.

MFDn is a platform-independent Fortran~90 code using a hybrid MPI+X programming
model.  Over the past decade, it has been successfully deployed on multi-core
supercomputers such as Jaguar at the Oak Ridge Leadership Class Facility (OLCF),
Mira at the Argonne Leadership Class Facility (ALCF), and Edison at the National
Energy Research Scientific Computing Center (NERSC) using MPI +
OpenMP~\cite{Maris:2011as,Maris:2013poa,Maris:2014jha,Binder:2015mbz}.
Currently it is in production on many-core
systems~\cite{10.1007/978-3-319-46079-6_26,LENPIC:2018ewt,Caprio:2021umc} such
as Cori at NERSC and Theta at ALCF, as well as other supercomputers worldwide.

As part of the NESAP application readiness program for NERSC’s  latest
Perlmutter system, MFDn is being updated to take advantage of accelerators.  The
current mainline GPU port uses OpenACC.  In this work we consider several
kernels that are representative for some of the most time-consuming parts of
MFDn \cite{10.5555/1413370.1413386,Maris_2013,10.1007/978-3-319-46079-6_26}.  We
describe some of the challenges and limitations of running them efficiently on
GPUs with OpenMP and/or OpenACC directives, using both the NVIDIA and Cray
Fortran compilers.  We test their performance on NVIDIA V100 ``Volta'', NVIDIA
A100 ``Ampere'', and AMD MI100 GPUs, as well as on Intel multi-core CPUs; see
Table~\ref{tab:hw_platformsw} for more details of the test systems.

\begin{table}[tb]
  \centering
  \caption{Test platforms}
  \label{tab:hw_platformsw}
  \begin{tabular}{|l|l|l|l|}
    \hline
                   & Cori GPU      & Cori DGX    & Spock     \\
    \hline
    GPU            & NVIDIA V100   & NVIDIA A100 & AMD MI100 \\
    CPU            & Intel Skylake & AMD Rome    & AMD Rome  \\
    GPUs per node  & 8             & 8           & 4         \\
    CPUs per node  & 2             & 2           & 1         \\
    Bus            & PCIe 3.0      & PCIe 4.0    & PCIe 4.0  \\
    per GPU Memory & 16GB          & 40 GB       & 32 GB     \\
    \hline
  \end{tabular}
\end{table}

We performed tests on Cori GPU (NVIDIA V100 GPUs) and Cori DGX (NVIDIA A100)
with the NVIDIA compiler only, whereas our tests on Spock (AMD MI100) are
performed with the Cray compiler only (see Table~\ref{tab:compilers}). The Cray
compiler version 12.0 was not available for the Cori GPU system at the time of
writing and it does not support NVIDIA A100 GPUs. All tests were done on
exclusively-allocated nodes. In order to minimize any NUMA effects, tests on
GPUs used a single GPU, while tests on CPUs used a single socket.

\begin{table}[tb]
  \centering
  \caption{Compilers. Cray Fortran compiler version 12.0.1 claims full OpenACC 2.0 and partial OpenACC 2.6 support for Fortran.}
  \label{tab:compilers}
  \begin{tabular}{|l|c|c|c|c|c|c|}
    \hline
    Vendor   & Version & \texttt{\_OPENACC} & \texttt{\_OPENMP} & V100       & A100       & MI100      \\
    \hline
    NVIDIA   & 21.7    & 201711 (2.6)       & 202011 (5.1)      & \checkmark & \checkmark &            \\
    HPE/Cray & 12.0.1  & 201306 (2.0)       & 201511 (4.5)      & \checkmark &            & \checkmark \\
    \hline
  \end{tabular}
\end{table}

In our implementations we have primarily explored three models: OpenACC with
\texttt{parallel} and \texttt{loop} directives, \ie no \texttt{kernels}
directives; OpenMP target offload with prescriptive style directives
(\texttt{teams distribute parallel do}); and OpenMP target with the new
\texttt{loop} construct introduced in OpenMP 5.0. In all cases maintaining
performance on CPUs and code maintainability was also a priority. The NVIDIA
compiler claims support for OpenACC 2.6 with ``many features'' from 2.7 and for
a subset of OpenMP 5.1, while the Cray Fortran compiler claims support for
OpenACC 2.0 with ``partial support'' for 2.6 and full support OpenMP 4.5 with
partial support for 5.0. We highlight where we encountered missing features or
shortcomings throughout and in particular in architectural specialization in
Section~\ref{section:fill_array} and reductions on arrays in
Section~\ref{section:array_reduction}.

\section{Computational Motifs in Configuration Interaction code MFDn}

The key computational challenges for MFDn are (1) efficient localization of the
nonzero Hamiltonian matrix elements and evaluation of the corresponding matrix
elements, and (2) efficient sparse matrix--vector and matrix--matrix products
used in the solution of the eigenvalue problem, both while effectively using the
available aggregate
memory~\cite{10.5555/1413370.1413386,Maris_2013,10.1007/978-3-319-46079-6_26}.
Figure~\ref{fig:MFDn_structure} shows the overall structure of MFDn.  The GPU
port of the LOBPCG eigensolver~\cite{SHAO20181} using OpenACC is described
in~\cite{Maris:2021mug}.  In this work we concentrate on the matrix construction
phase and the evaluation of observables, which each take up about one-quarter to
one-third of the total runtime, while iterative eigensolving takes about
one-third to one-half of the total runtime.
\newsavebox{\MFDnstructure}
\begin{lrbox}{\MFDnstructure}
  \begin{minipage}{0.95\textwidth}
    \begin{enumerate}
      \item Enumerate and distribute many-body basis orbitals ($\Psi_j$)
      \item \label{tile_cnt}
            Loop over column orbitals ($\Psi_m$)\\
            \mbox{}\quad Loop over row orbitals ($\Psi_l$)
      \item \mbox{}\quad\quad Compare the single-particle orbitals of $\Psi_l$ and $\Psi_m$\\
            \mbox{}\quad\quad increment $tile\_cnt$ if up to $2d$ differences\\
            \mbox{}\quad End Loop\\
            End Loop
      \item Allocate arrays of length $tile\_cnt$ to store nonzero tiles ($T_k$)
      \item Repeat \ref{tile_cnt} and store nonzero tiles ($T_k = (\Psi_l,\Psi_m)$):\\
            pairs of orbitals with up to $2d$ differences
      \item Loop over tiles ($(\Psi_l,\Psi_m) = T_k$)
      \item \mbox{}\quad Loop over column states ($\Phi_j \in \Psi_m$)\\
            \mbox{}\quad\quad Loop over row states ($\Phi_i \in \Psi_l$)
      \item \mbox{}\quad\quad\quad Compare the single-particle states of $\Phi_i$ and $\Phi_j$\\
            \mbox{}\quad\quad\quad increment $cnt_k$ if up to $2d$ differences\\
            \mbox{}\quad\quad End Loop\\
            \mbox{}\quad End Loop\\
            End Loop
      \item Convert $cnt_k$'s to {\it offset}$_k$'s
      \item Allocate arrays of length $\sum cnt_k$ to store $H_{ij}$
      \item Loop over tiles ($(\Psi_l,\Psi_m) = T_k$)
      \item \mbox{}\quad Loop over column states ($\Phi_j \in \Psi_m$)\\
            \mbox{}\quad\quad Loop over row states ($\Phi_i \in \Psi_l$)
      \item \mbox{}\quad\quad\quad Compare the single-particle states of $\Phi_i$ and $\Phi_j$\\
            \mbox{}\quad\quad\quad cycle if more than $2d$ differences
      \item \mbox{}\quad\quad\quad Compute the nonzero matrix entry $H_{ij}$ and store\\
            \mbox{}\quad\quad End Loop\\
            \mbox{}\quad End Loop\\
            End Loop
      \item Obtain lowest $n$ eigenvalues $E$ and eigenvectors $\vec{c}$ of $H_{ij}$\\
            using distributed LOBPCG or Lanczos algorithm
      \item Loop over tiles ($(\Psi_l,\Psi_m) = T_k$)
      \item \mbox{}\quad Loop over column states ($\Phi_j \in \Psi_m$)\\
            \mbox{}\quad\quad Loop over row states ($\Phi_i \in \Psi_l$)
      \item \mbox{}\quad\quad\quad Compare the single-particle states of $\Phi_i$ and $\Phi_j$\\
            \mbox{}\quad\quad\quad cycle if more than $2d$ differences
      \item \mbox{}\quad\quad\quad Compute $m$ nonzero matrix elements $O_{ij}(1:m)$
      \item \mbox{}\quad\quad\quad Update $a(1:n*m) = a(1:n*m) + c_i(1:n) * O_{ij}(1:m) * c_j(1:n)$\\
            \mbox{}\quad\quad End Loop\\
            \mbox{}\quad End Loop\\
            End Loop
    \end{enumerate}
  \end{minipage}
\end{lrbox}
\begin{figure}
  \fbox{\usebox{\MFDnstructure}}
  \caption{Schematic outline of the structure of MFDn during the matrix construction phase (lines 2 through 14) and evaluation of physical observables (lines 16 through 20), for a $d$-body Hamiltonian and $d$-body operators for observables.}

  \label{fig:MFDn_structure}
\end{figure}

In CI calculations, the many-body wave functions are approximated by an
expansion in {\em many-body basis states}; in MFDn we use antisymmetrized
products of single-particle states with quantum numbers $(n, \ell, j, m)$ (see,
e.g., Ref.~\cite{Suhonen:2007vjh} for the meaning of these quantum numbers). The
many-body basis states can then be characterized by the set of single-particle
quantum numbers for each nucleon.  It is convenient to group together many-body
states with the same sets of values for the quantum numbers $(n, \ell, j)$, but
different sets of values for the magnetic projection quantum numbers $m$. This
grouping leads to a natural hierarchy in the sparsity structure of the
Hamiltonian matrix, which in turn allows for efficient localization and
evaluation of nonzero matrix elements.  In addition, this grouping also
facilitates an efficient block-diagonal preconditioner for the LOBPCG
algorithm~\cite{SHAO20181}.  We refer to these groups of many-body states as
{\em many-body basis orbitals}.  To describe the sparsity structure, we
furthermore define tiles as pairs of row and column many-body basis orbitals.

In order to efficiently locate the nonzero matrix elements we exploit this
hierarchical structure, by first determining which tiles can contain nonzero
matrix elements  (lines 2 through 5 of Fig.~\ref{fig:MFDn_structure}) and next
counting how many nonzero matrix elements there are (lines 6 through 8) in each
tile.  The actual construction of the sparse matrix starts in line 11, with the
nonzero matrix elements and their location evaluated and stored in line 14 of
Fig.~\ref{fig:MFDn_structure}.  Note that the structure of the double loops
starting in lines 2 and in line 7 is essentially the same; the corresponding
motif is discussed in Section~\ref{section:sparsity} below.  Also the double
loops starting in line 12 have a similar structure, except that in this case,
the obtained results in the innermost loop are stored in an array; this motif is
discussed in Section~\ref{section:fill_array} below.  (Note that this motif is
also used implicitly in line 5 of Fig.~\ref{fig:MFDn_structure}.)

Once the nonzero matrix elements are evaluated and stored, we can obtain the
lowest $n$ eigenvalues and eigenvectors $\vec{c}$ using an iterative
eigensolver. In MFDn we use either LOBPCG or a Lanczos algorithm with
reorthogonalization, as indicated in line 15 of Fig.~\ref{fig:MFDn_structure}.
Details of the GPU port of our LOBPCG eigensolver using OpenACC are described in
Ref.~\cite{Maris:2021mug}.

Finally, in lines 16 through 20 of Fig.~\ref{fig:MFDn_structure} we calculate
$m$ different physical observables using the coefficients $c_i$ of the lowest
$n$ eigenvectors and $m$ two-body operators.  Typically, we use $8$ or $16$
eigenvectors, and up to $m \sim 16$ different operators corresponding to
different observables.  Note that lines 16 to 18 have the same structure as
lines 11 to 13, but instead of storing the nonzero matrix elements of the
operators, we contract them with the $n$ eigenvectors.  The corresponding motif
for these loops is discussed in Section~\ref{section:array_reduction} below.

\subsection{Matrix Sparsity Determination}
\label{section:sparsity}

A typical loop structure in the matrix construction phase, as well as in the
evaluation of observables, is shown in Fig.~\ref{fig:loop}.  The $(i,j)$th entry
of the many-body Hamiltonian with a $d$-body interaction, $H_{ij} = \langle
\Phi_i \vert H \vert \Phi_j \rangle$, can only be nonzero if the many-body
states $\Phi_i$ and $\Phi_j$ differ by at most $2d$ single-particle states.
Thus, the first step in the matrix construction (and in the evaluation of
observables given by the expectation value of a $d$-body operator) is to
determine which matrix elements can be nonzero.  This is done in line 3 of the
loop; subsequently, lines 4 and 5 indicate the actual evaluation of the nonzero
matrix entry, which can be stored in memory (see line 14 of
Fig.~\ref{fig:MFDn_structure}) , or directly used in a matrix--vector multiply
or vector--matrix--vector contraction for the calculation of expectation values
(see line 20 of Fig.~\ref{fig:MFDn_structure}).  These are more complicated
operations which are accomplished by separate (sequential) subroutine calls, the
details of which are beyond the scope of this work.

\newsavebox{\loopsimple}
\begin{lrbox}{\loopsimple}
  \begin{minipage}{0.95\textwidth}
    \begin{enumerate}
      \item Loop over column states ($\Phi_j$)
      \item \mbox{}\quad Loop over row states ($\Phi_i$)
      \item \mbox{}\quad\quad Compare the single-particle states of $\Phi_i$ and $\Phi_j$\\
            \mbox{}\quad\quad cycle if more than $2d$ differences
      \item \mbox{}\quad\quad (optional) Compute the nonzero matrix entry $H_{ij}$ and store
      \item \mbox{}\quad\quad (optional) FMA of $H_{ij}$ with $i$th row (and $j$th column) vector element
      \item \mbox{}\quad End Loop
      \item End Loop
    \end{enumerate}
  \end{minipage}
\end{lrbox}
\begin{figure}
  \fbox{\usebox{\loopsimple}}
  \caption{A typical loop structure in the matrix construction phase and during the evaluation of physical observables.}
  \label{fig:loop}
\end{figure}

The localization of nonzero matrix elements involves determining which many-body
basis states may be connected by the given particle rank (\eg two-body)
Hamiltonian. Given a single-particle basis with $n_\mathrm{s.p.}$
single-particle states, a many-body basis state $\Phi_i$ for fermionic systems
can be represented by a binary string of length $n_\mathrm{s.p.}$, denoted by
$\BIN(\Phi_i)$, where each binary bit of $\BIN(\Phi_i)$ indicates whether the
corresponding single-particle state is occupied. The total number of particles
in the many-body state $\Phi_i$ is the number of 1's in $\BIN(\Phi_i)$, \ie the
population count $\mathtt{popcount}(\BIN(\Phi_i))$. Information regarding all
differently-occupied single-particle states between two bit-representations
$\BIN(\Phi_i)$ and $\BIN(\Phi_j)$ is encoded by $\BIN(\Phi_i)\XOR\BIN(\Phi_j)$,
where $\XOR$ denotes the bit-wise exclusive-or operation.  The number of
differently-occupied single-particle states is then
$\mathtt{popcount}(\BIN(\Phi_i)\XOR\BIN(\Phi_j))$.  If both bit-representations
describe states with the same number of particles, then the number of
differently-occupied single-particle states is always even; with a two-body
potential, only many-body matrix elements with 0, 2, or 4 differently-occupied
single-particle states can be nonzero. If there are more than four
differently-occupied single-particle states the matrix element must be zero.

The storage requirements of such a bit representation is proportional to the
number of single-particle states, but independent of the number of particles in
the system.  Alternatively, one can represent an $N$-body basis state $\Phi_i$
by an array or tuple of $N$ short integers $\MBS(a_1{:}a_N)$ with $a_1 < a_2 <
\cdots < a_N$, where each element $a_i$ indicates which single-particle state is
occupied.  The storage requirement of this method is proportional to $N$, the
number of particles, but independent of the number of single-particle states
$n_\mathrm{s.p.}$.  For a relatively small number of particles ($N\sim 10-20$ in
MFDn), but a large ($n_\mathrm{s.p.} \gtrsim 1,000$) number of single-particle
states, the $\MBS$ representation is more advantageous in terms of memory than
storing the states as a bit representation.  However, determining the
differently-occupied single-particle states is significantly more complex when
$\Phi_i$ and $\Phi_j$ are in the $\MBS$ representation (see Listing
\ref{listing:cnt_difference}).  Note the factor of two to ensure consistent
counts with the population count on the bit representation.
\begin{listing}
  \begin{minted}{fortran}
integer function count_difference(s1, n1, s2, n2)
  integer, intent(in) :: n1, n2, s1(n1), s2(n2)
  integer :: i1, i2, d, diffs1, diffs2
  !$acc routine seq
  i1 = 1
  i2 = 1
  diffs1 = 0
  diffs2 = 0
  do
     if ( (i1 > n1) .or. (i2 > n2) ) exit
     d = s1(i1) - s2(i2)
     if (d < 0) then
        diffs1 = diffs1 + 1
        i1 = i1 + 1
     else if (d > 0) then
        diffs2 = diffs2 + 1
        i2 = i2 + 1
     else
        i1 = i1 + 1
        i2 = i2 + 1
     end if
  end do
  count_difference = 2*max(diffs1, diffs2)
end function count_difference
\end{minted}
  \caption{Sequential function for detailed comparison of two many-body states.}
  \label{listing:cnt_difference}
\end{listing}

In MFDn the low-lying single-particle states are most likely to be occupied.
For this reason we use a bit representation for the 64 lowest single-particle
states, in combination with an array of $N$ 16-bit integers $\MBS(a_1{:}a_N)$ to
store the full many-body state.  This allows for efficient filtering of pairs of
states with bit arithmetic on the most likely to be occupied states while
incurring minimal additional storage overhead.  Our OpenACC implementation of
the algorithm used to count the number of different single-particle states by
first performing a population count on the 64 lowest single-particle states,
followed by a detailed comparison of the $\MBS$ representation if the population
count is at most $2d$, is given in code Listings \ref{listing:cnt_difference}
and \ref{listing:count_nonzero}.
\begin{listing}
  \begin{minted}[highlightlines={1,4}]{fortran}
!$acc parallel loop
do i = 1, n
   c = 0
   !$acc loop reduction(+:c)
   do j = 1, n
      d = popcnt(ieor(bitrep1(i), bitrep2(j)))
      if (d > 4) cycle
      d = count_difference(mbstate1(:,i), np, mbstate2(:,j), np)
      if (d <= 4) c = c + 1
   end do
   !$acc end loop
   counts(i) = c
end do
!$acc end parallel loop
numnnz = sum(counts)
\end{minted}
  \caption{Counting nonzero matrix elements with OpenACC, using both a bit representation for the first 64 single-particle states of $\Phi_i$ and $\Phi_j$ and a detailed comparison for the $\MBS$ representation if needed. The highlighted lines show the directives used for the two levels of parallelism.}
  \label{listing:count_nonzero}
\end{listing}

OpenMP prescriptive, OpenMP loop and OpenMP loop with bind hints directives for
the first and second  level of parallelism in Listing
\ref{listing:count_nonzero} are shown in Listings \ref{listing:count_level1} and
\ref{listing:count_level2} respectively. The first level is typically mapped to
cores on CPUs, thread blocks on NVIDIA GPUs and workgroups on AMD GPUs while the
second level is typically mapped to SIMD lane(s) on CPUs, threads on NVIDIA GPUs
and work items on AMD GPUs. Note that the comparison of the two complete
many-body states is performed in a function (see Listing
\ref{listing:cnt_difference}), which needs a \texttt{!\$acc routine seq}
directive so that the OpenACC loops in Listing \ref{listing:count_nonzero} do
indeed get parallelized; with OpenMP there is no need for a similar directive
when the routine is defined in the same compilation unit, though \texttt{!\$omp
declare target} may be used when this is not the case.

\begin{listing}
  \begin{minted}{fortran}
!$acc parallel loop
!$omp target teams distribute private(d)
!$omp target teams loop private(d)
!$omp target teams loop bind(teams) private(d)
\end{minted}
  \caption{OpenACC, OpenMP prescriptive, OpenMP loop and OpenMP loop with hints directives to express the first level of parallelism highlighted in line 1 of Listing \ref{listing:count_nonzero}.}
  \label{listing:count_level1}
\end{listing}

\begin{listing}
  \begin{minted}{fortran}
!$acc loop reduction(+:c)
!$omp parallel do reduction(+:c) private(d)
!$omp loop reduction(+:c) private(d)
!$omp loop bind(parallel) reduction(+:c) private(d)
\end{minted}
  \caption{OpenACC, OpenMP prescriptive, OpenMP loop and OpenMP loop with hints directives to express the second level of parallelism highlighted in line 4 of Listing \ref{listing:count_nonzero}.}
  \label{listing:count_level2}
\end{listing}

\begin{figure}
  \includegraphics[width=\textwidth]{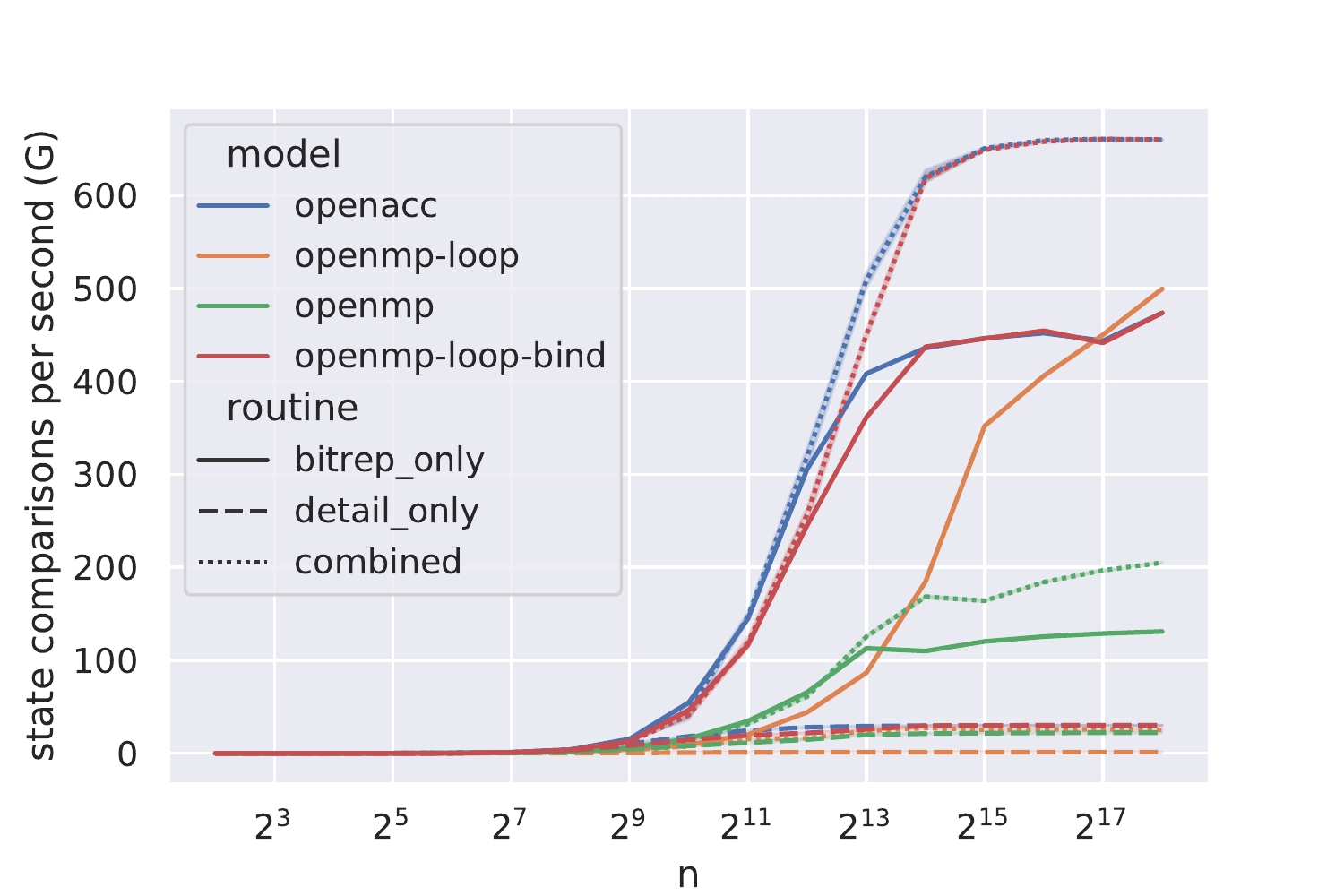}
  \caption{Performance of interacting state counting routines on A100 with 8
  particles (median density of nonzero elements is  $\sim 6\times 10^{-6}$). The
  vertical axis shows the number of state comparisons made per second,
  $\textrm{rate} = n^2 / \textrm{time}$ (higher is better).}
  \label{fig:count_a100}
\end{figure}

\begin{figure}
  \centering
  \includegraphics[width=\textwidth]{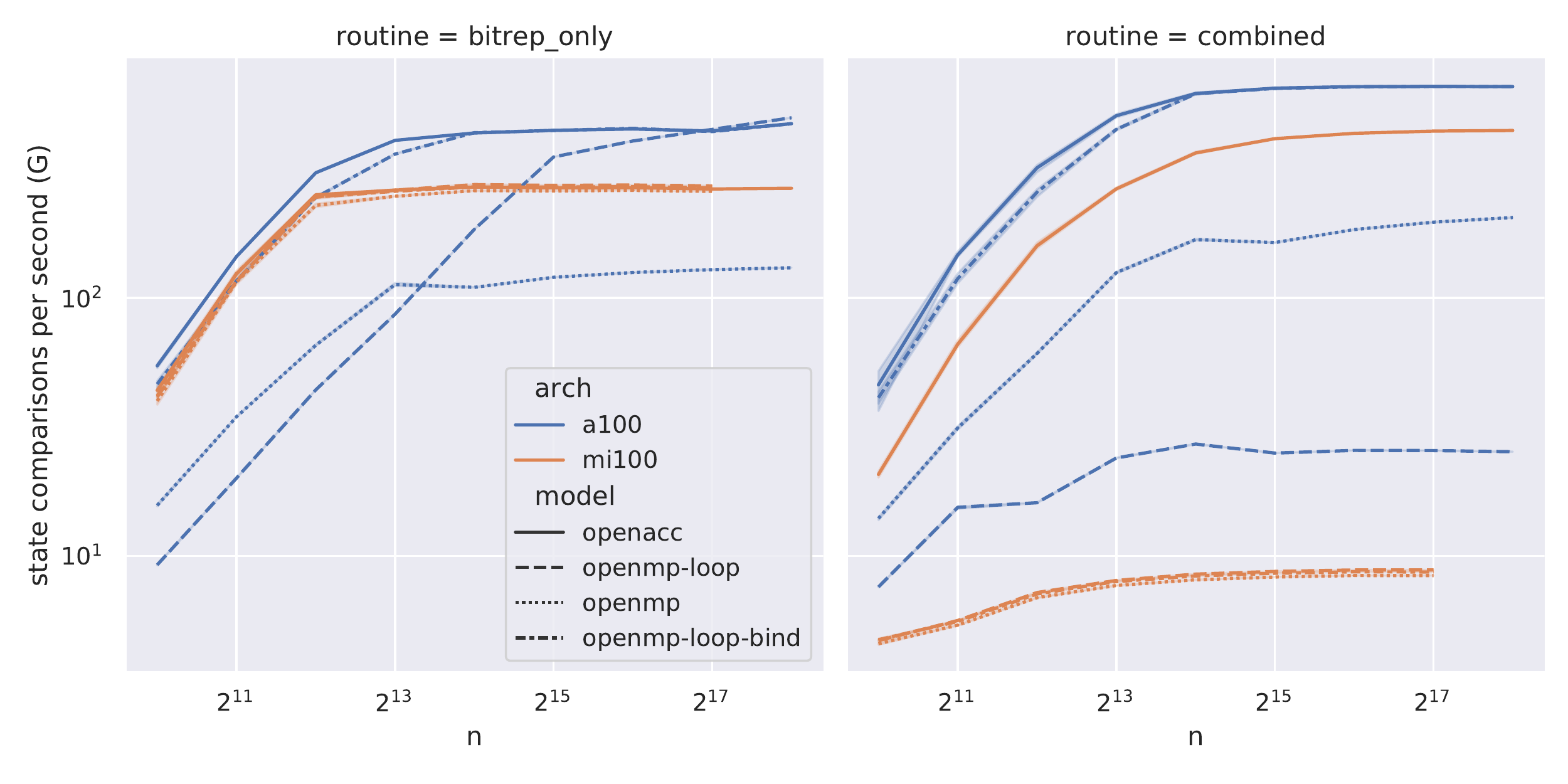}
  \caption{Performance of bit representation only and combined interacting state
  counting routines on A100 and MI100 with 8 particles (median density of
  nonzero elements is  $0.000006$). The vertical axis shows the number of state
  comparisons made per second, $\textrm{rate} = n^2 / \textrm{time}$ (higher is
  better).}
  \label{fig:count_arch}
\end{figure}

For our performance tests we used many-body states with 4, 8, 12, 16 and 20
particles, 128 single-particle states, and a bit representation based on only
the lowest 64 single-particle states.  We randomly generated many-body states
biased towards the lowest states, and counted the number of nonzero matrix
elements for a two-body interaction. We present in Figs.~\ref{fig:count_a100}
and \ref{fig:count_arch} results with 8 particles as the density of nonzeros
(median density was $6\times 10^{-6}$) most closely represents the regime of
interest for MFDn.  Note that fewer particles with the state generation scheme
correspond to more nonzero elements and more particles result in fewer nonzero
elements. For comparison and correctness, we also ran two additional versions: a
version without the \texttt{popcount} on the bit representation, as well as a
version with a complete bit representation of all 128 single-particle states and
using exclusively the \texttt{popcount} on this extended bit representation.

Figure~\ref{fig:count_arch} shows the performance of the bit representation and
combined  versions of the counting routines on MI100 and A100 GPUs implemented
with OpenACC and OpenMP. On NVIDIA and AMD GPUs the OpenACC directives provide
the best performance in most cases. In several instances there were performance
issues when a function or subroutine call was introduced: particularly with
OpenMP, usually manifest as a failure to generate parallel code for the second
level of parallelism. In the case of \texttt{!\$omp loop} directives we found
that it was necessary to include bind annotations to recover the performance
obtained by the OpenACC implementations on NVIDIA GPUs. The performance
difference between all versions and implementations is shown in
Fig.~\ref{fig:count_a100}. On NVIDIA platforms we found there is overhead
associated with  OpenMP with prescriptive \texttt{!\$omp target teams
distribute} and \texttt{!\$omp parallel do} directives compared to the
\texttt{!\$omp loop} directives. The compiler parallelizes on both teams and
threads, but shows reduced performance.  This is likely due to the additional
semantic constraints for the \texttt{!\$omp parallel} directive introducing some
overhead in code generated by the compiler.

\subsection{Parallel prefix sum}
\label{section:prefix_sum}
On multi-core CPUs, it is often convenient to use private arrays -- and as long
as there is sufficient memory, there is no intrinsic limitation on the private
array size. In practice, it is limited by the \texttt{OMP\_STACKSIZE} which the
user can increase from its default value if necessary. Further, the use of
thread-private arrays can often result in good performance as it helps ensure
cache locality on CPUs.  However, the situation on GPUs is different.  Although
private arrays can be used in both OpenACC and with OpenMP offload, there are
more limitations on the size of such private arrays and/or on the number of
gangs/teams or vector length one can use, due to the order of magnitude more
parallelism available in GPUs.  In particular in inner loops, private arrays
should be avoided or limited to small arrays with only a handful of array
elements; and even at the gang/team level, large arrays with (tens of) thousands
of array elements severely limits the number of gangs/teams one can use.

In order to reduce (or better, completely avoid) the need for private arrays, it
can be useful to convert counts, such as the counts of nonzero matrix elements
discussed in Section \ref{section:sparsity}, into offsets, so that one can use a
single  large shared array with appropriate offsets, instead of many allocatable
private arrays.  Specifically, converting counts $x_i$ into offsets $y_i$ can be
implemented as
\begin{equation}
  y_{i+1} = \sum_{j=0}^{i} x_j = y_i + x_i
\end{equation}
with $y_1=0$, and is often referred to as a prefix sum or cumulative sum or
scan. Here we focus on addition of integers, but in general only a binary
associative operator is required.

\begin{listing}[b]
  \begin{minted}{fortran}
!$acc serial present(x,y)
y(1) = 0
do i = 1, n-1
   y(i+1) = y(i) + x(i)
end do
!$acc end serial
\end{minted}
  \caption{Serial prefix sum with OpenACC.}
  \label{listing:scan_serial_acc}
\end{listing}
On CPUs this operation is fast, and furthermore OpenMP 5.0 introduced a
\texttt{scan} directive that extends reductions. However, this feature is not
currently supported for GPUs by the compilers tested in this work. OpenACC
provides no equivalent directive. We note that production quality
implementations of this operation may be available in C++ libraries such as
Thrust~\cite{CarterEdwards20143202} or Kokkos~\cite{bell2012thrust}, but that
including C++ or vendor specific frameworks in a Fortran code with a goal of
portability introduces significant maintenance costs. In many cases it is
preferable to avoid data transfers between the host and accelerator, even at the
cost of inefficient use of the device. With OpenACC's \texttt{!\$acc serial}
directive a potentially expensive data transfer can be avoided as shown in
Listing~\ref{listing:scan_serial_acc}. Generally one should consider a
performance model that includes bandwidth between host and device, performance
on either host and device and the potential for any latency/ blocking effects
introduced by the data motion when considering an implementation.

For prefix sums over large sequences, a parallel implementation can realize
significant speedups. We note that in MFDn the offsets can be reused many times
so that the overall impact for the application run time is small, but this
common primitive can be found in many applications~\cite{BlellochTR90}. A
work-efficient algorithm for a parallel scan is shown in Algorithm
\ref{algo:parallel_prefix} - the work-efficient algorithm consists of an up and
down sweep~\cite{harris2007parallel, BlellochTR90}. The main idea is to sweep up
and down a binary tree of the input data. The ``up'' or ``reduce'' sweep
proceeds from the leaves to the root, computing partial sums in place. In the
``down'' sweep phase the binary tree is traversed from the last element down
(root) to the leaves.

\begin{algorithm}[t]
  \caption{Work-efficient parallel prefix sum (Algos. 3 and 4 of Ref.~\cite{harris2007parallel})}
  \label{algo:parallel_prefix}
  \begin{algorithmic}[1]
    \For {$p \gets 0, \log_2{n-1}$} \Comment{sweep up (or reduce)}
    \For {$j \gets 0,n-1$ by $2^{p+1}$} \Comment{parallel}
    \State   $y\left(j+2^{p+1}\right) \gets y\left(j+2^{p}\right) + y\left(j+2^{p+1}\right)$
    \EndFor
    \EndFor
    \State $x(n) \gets 0$
    \For {$p \gets \log_2{n}$,0} \Comment{sweep down}
    \For {$j \gets 0,n-1$ by $2^{p+1}$} \Comment{parallel}
    \State $tmp \gets y\left(j+2^p\right)$
    \State $y\left(j+2^p\right) \gets y\left(j+2^{p+1}\right)$
    \State $y\left(j+2^{p+1}\right) \gets tmp + y\left(j+2^{p+1}\right)$
    \EndFor
    \EndFor
  \end{algorithmic}
\end{algorithm}

\begin{listing}
  \begin{minted}{fortran}
!$acc data present(x,y)
!$acc parallel loop async
do j = 1, n
   y(j) = x(j)
end do
!$acc end parallel
offset = 1
! sweep up, reduction in place
do while (offset < n)
   !$acc parallel loop firstprivate(offset) present(y) async
   do concurrent (j=0:n-1:2*offset)
      y(j + 2*offset) = y(j + offset) + y(j + 2*offset)
   end do
   !$acc end parallel
   offset = 2*offset
end do
! sweep down, complete the scan
!$acc serial async
y(n) = 0
!$acc end serial
offset = rshift(offset, 1)
do while(offset > 0)
   !$acc parallel loop firstprivate(offset, tmp) present(y) async
   do concurrent(j=0:n-1:2*offset) local(tmp)
      tmp = y(j + offset)
      y(j + offset) = y(j + 2*offset)
      y(j + 2*offset) = tmp + y(j + 2*offset)
   end do
   !$acc end parallel
   offset = rshift(offset, 1)
end do
!$acc wait
!$acc end data
\end{minted}
  \caption{Parallel scan with OpenACC corresponding to Algorithm~\ref{algo:parallel_prefix}}
  \label{listing:scan_acc}
\end{listing}

\begin{figure}[t]
  \centering
  \includegraphics[width=0.90\textwidth]{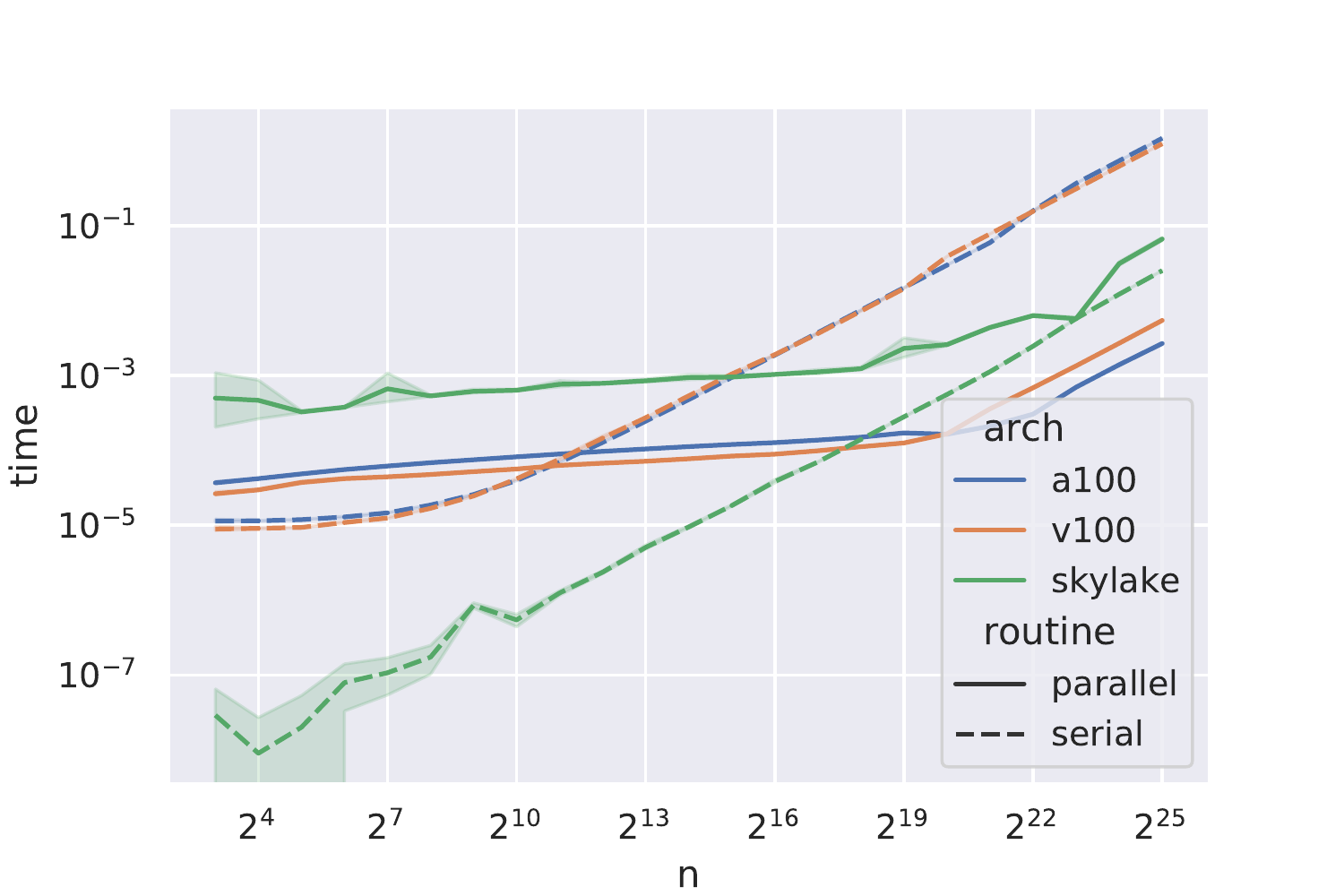}
  \caption{Performance of prefix sum implemented with OpenACC on different
  architectures in parallel and serial (lower is better).}
  \label{fig:scan_arch_perf}
\end{figure}

An implementation of Algorithm \ref{algo:parallel_prefix} in OpenACC is shown in
Listing~\ref{listing:scan_acc}. As written it assumes power of two arrays;
non-power of two arrays can be padded with zeros. We note that there are many
possible further optimizations and refer interested readers to
Refs.~\cite{harris2007parallel, BlellochTR90}. In OpenACC, each
\texttt{parallel} region will result in a new kernel launch, but with the
\texttt{async} clause we can queue them all in non-blocking manner and rely that
they will be executed in order. A similar approach can be implemented in OpenMP
with \texttt{nowait} and \texttt{depend} clauses.
Figure~\ref{fig:scan_arch_perf} shows the performance of serial and parallel
prefix sum with OpenACC on A100, V100, and Skylake. Unfortunately the Cray
compiler's partial support for OpenACC 2.6 does not include the \texttt{serial}
directive so we do not include MI100 results for these implementations. For
large arrays on A100 GPUs the parallel implementation can be over 500x faster
than a serial implementation. This demonstrates the importance of support for
serial work on accelerators (and corresponding constructs) for productivity,
support for asynchronous work for performance and the desirability of a good
language/ library support for parallel primitives such as prefix sums.

\subsection{Filling shared arrays}
\label{section:fill_array}
After an initial pass to obtain the nonzero counts and offsets as described in
Sections~\ref{section:sparsity} and \ref{section:prefix_sum} a second pass is
often performed to store relevant information such as a row or column index, or
the nonzero value, into a global (shared) array, see e.g. line 14 of
Fig.~\ref{fig:MFDn_structure}.  Since we are using a multilevel hierarchical
structure for the sparse matrix, this motif appears in several situations, not
only for the matrix elements themselves. Generally there are two levels of
parallelism: an outer loop, with no data dependencies, and an inner loop, where
order does not matter but there is a dependency. The outer loop alone typically
has enough parallelism to saturate a CPU but not a GPU.

\begin{figure}
  \includegraphics[width=0.90\textwidth]{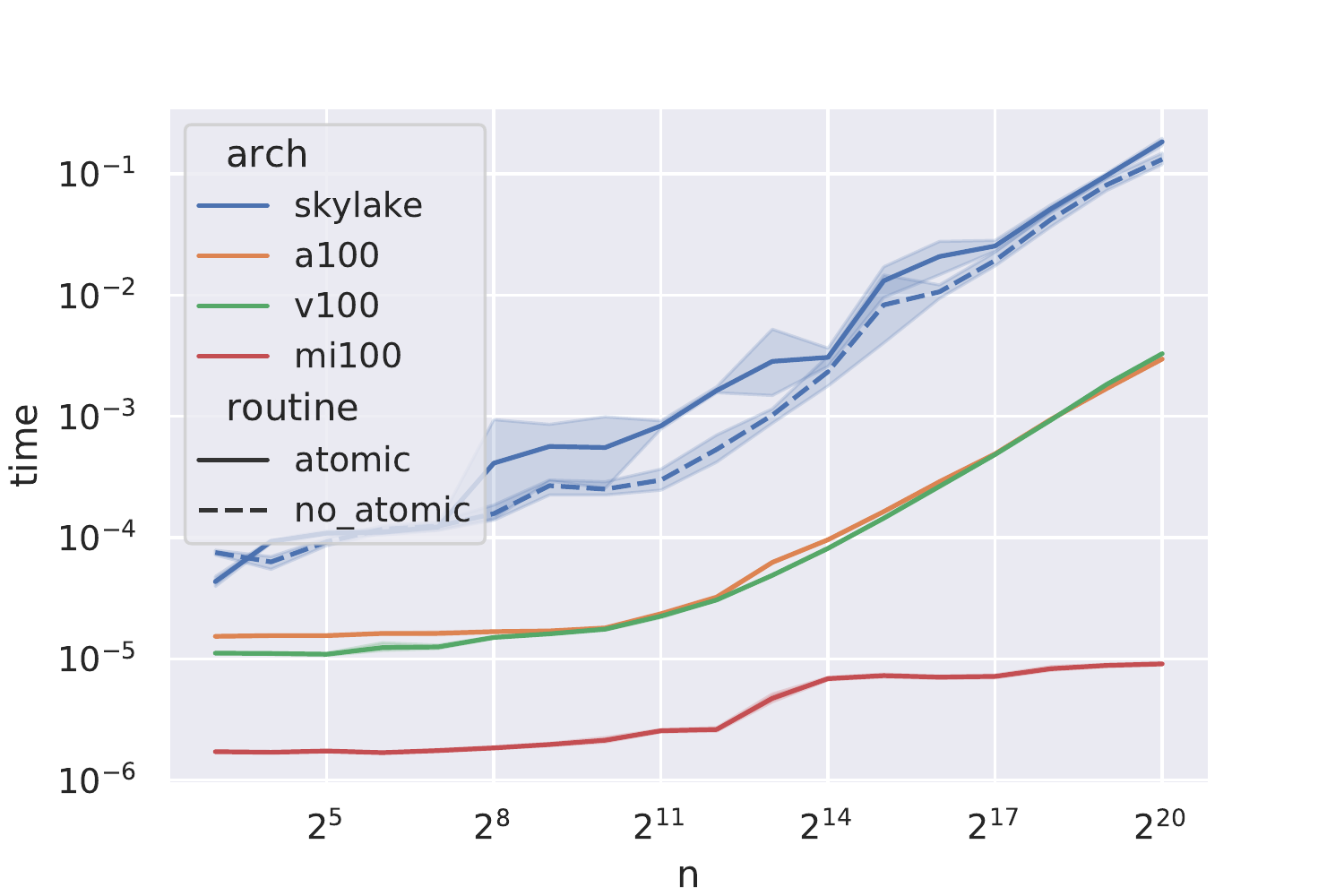}
  \caption{Performance of filling a shared array with OpenACC
  (Listing~\ref{listing:filling}) with $m=512$ and $p=1$ on MI100, A100, V100
  and Skylake (lower is better).}
  \label{fig:fill-array-by-arch}
\end{figure}

On GPUs, OpenACC directives can be used to efficiently implement such a motif as
shown in Listing \ref{listing:filling}. Equivalent directives are available in
OpenMP. The \texttt{!\$acc atomic capture} directive ensures that the value of
the shared array element \mintinline{fortran}{indx(i)} gets incremented by one
and assigned to the local (private) variable \mintinline{fortran}{k}, which can
then be used as an index for filling the desired array with the appropriate
value.  On GPUs the performance penalty attributed to atomic operation in the
inner loop is rather modest, and the exposed parallelism of the inner loop
overwhelms this penalty, resulting in significant speedup over CPUs as shown in
Fig.~\ref{fig:fill-array-by-arch}.

\begin{listing}
  \begin{minted}{fortran}
!$acc parallel loop
do i = 1, n
   indx(i) = offset(i)
   !$acc loop device_type(host) seq
   do j = 1, m
      if (mod(j,p) == 0) then
         !$acc atomic capture
         indx(i) = indx(i) + 1
         k = indx(i)
         !$acc end atomic
         arr(k) = j
      end if
   end do
end do
!$acc end parallel
\end{minted}
  \caption{Filling shared arrays on GPUs using OpenACC.}
  \label{listing:filling}
\end{listing}

The same source code can also be compiled for and run on CPUs. On CPUs, when the
parallelism available in the outermost level is sufficient, we can indicate that
the inner loop should be sequential with the addition of the \texttt{!\$acc
device\_type(host) seq} clauses.  Unfortunately, the OpenACC specification does
not support the \texttt{!\$acc device\_type} clause on \texttt{!\$acc atomic}
constructs. The performance of Listing~\ref{listing:filling} on multiple
architectures is shown in Figure~\ref{fig:fill-array-by-arch}. With the atomic
operation explicitly in the inner loop, this implementation performs worse on
CPUs than necessary despite there being no contention between threads on the
atomic operations due to the overhead of atomic semantics. In OpenMP 5.0
\texttt{metadirective} was introduced to support this use case, but compiler
support is not available at the time of writing. In principle, \texttt{declare
variant} in OpenMP or runtime calls in OpenACC to selectively choose between
multiple subroutine versions could be used to enable performance on CPUs and
GPUs with a single source at the expense of code duplication. Otherwise use of
the preprocessor would be required.

\subsection{Array Reductions}
\label{section:array_reduction}
Finally, to compute physical observables one needs to calculate the expectation
values of the corresponding operators $O_k$ (see line 20 of
Fig.~\ref{fig:MFDn_structure}):
\begin{equation}
  a_k = \sum_{ij} x_i\, (O_k)_{ij} \; y_j \;.
\end{equation}
In practice this requires a reduction of an array with a small number of
elements. In the context of the MFDn application the array is typically of
dimension $m \in [8,256]$. Here we consider a simplified version of the motif
(shown in Listing~\ref{listing:array_reduction_array}) that omits the check if
two many-body states interact and any computation of additional physical matrix
elements in order to explore programming model support for the motif. In this
application trading memory for performance as in \cite{kim2021gpu} is
undesirable since it will limit scaling to large problems on full systems. We
consider 3 implementations for ``small'' array reductions:
\begin{itemize}
  \item reduction clauses with array arguments;
  \item atomic updates to individual array elements;
  \item generating a scalar reduction for each element of the array with \texttt{fypp}.
\end{itemize}

\begin{listing}
  \begin{minted}{fortran}
!$acc parallel loop collapse(2) reduction(+:a)
do i = 1, n
   do j = 1, n
      do k = 1, m
         a(k) = a(k) + x(k,i) * y(k,j)
      end do
   end do
end do
!$acc end parallel
\end{minted}
  \caption{Array reduction: array variable in a \texttt{reduction} clause.}
  \label{listing:array_reduction_array}
\end{listing}

\subsubsection{direct array reduction support}
\begin{figure}
  \centering
  \includegraphics[width=0.7\textwidth]{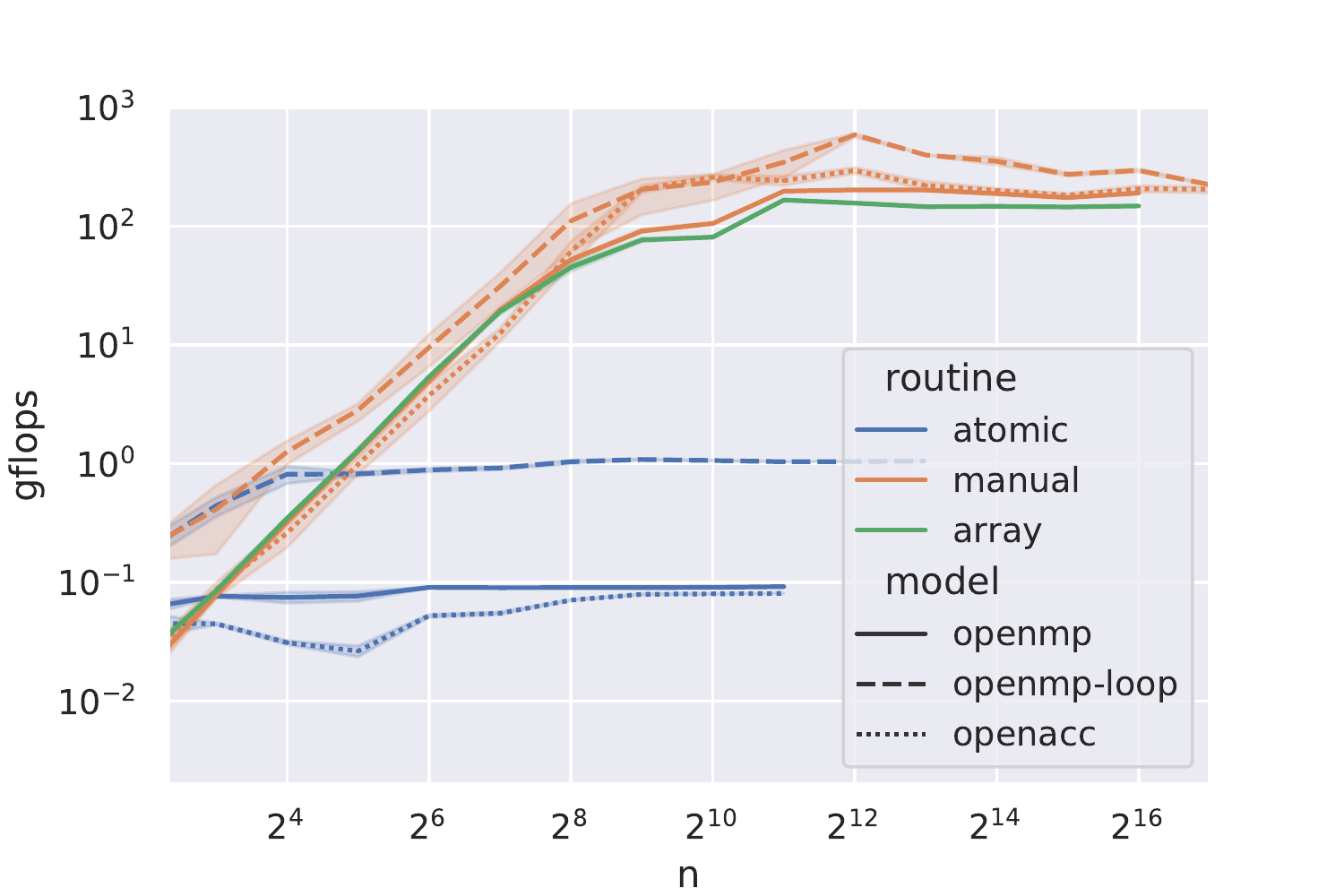}
  \includegraphics[width=0.7\textwidth]{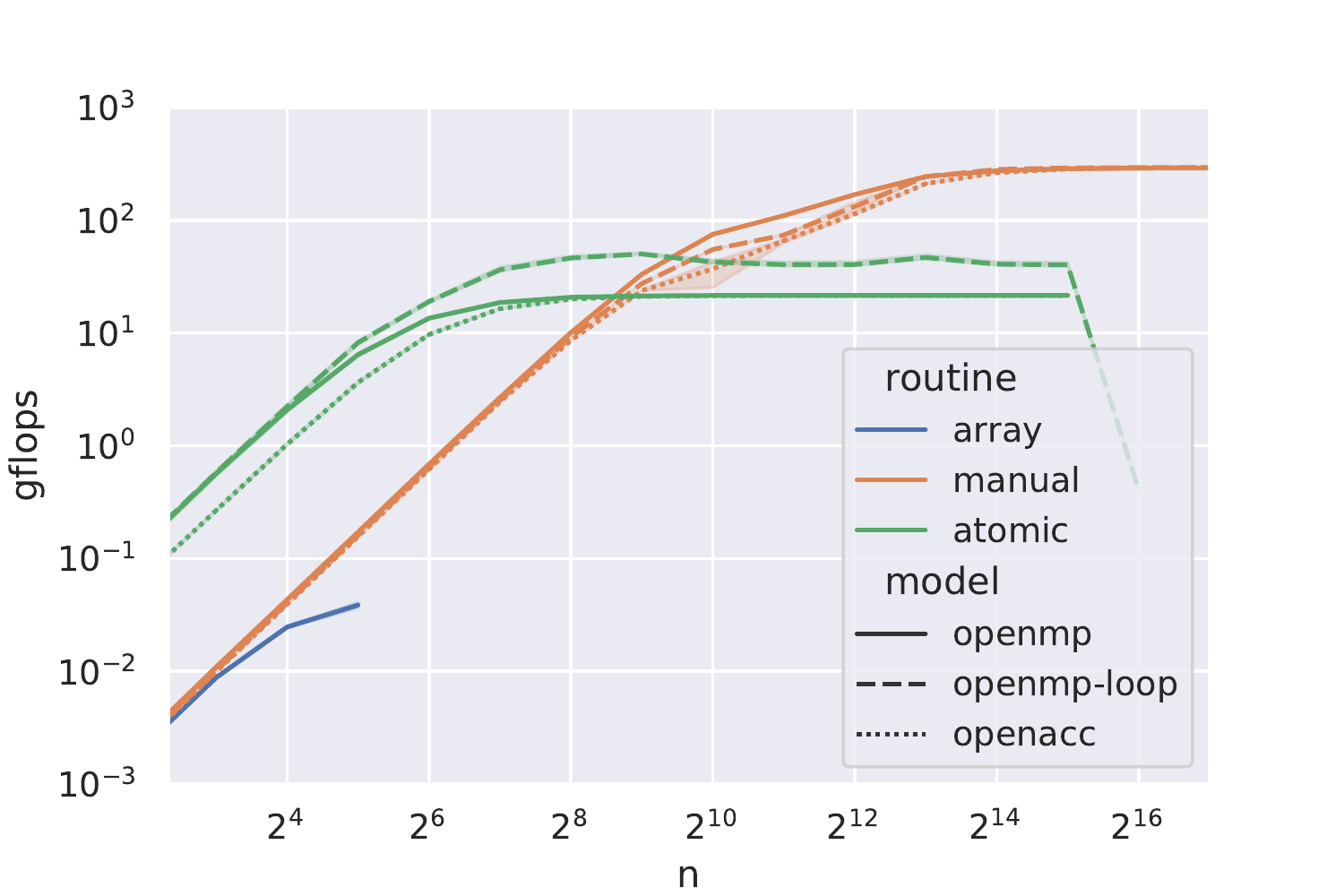}
  \includegraphics[width=0.7\textwidth]{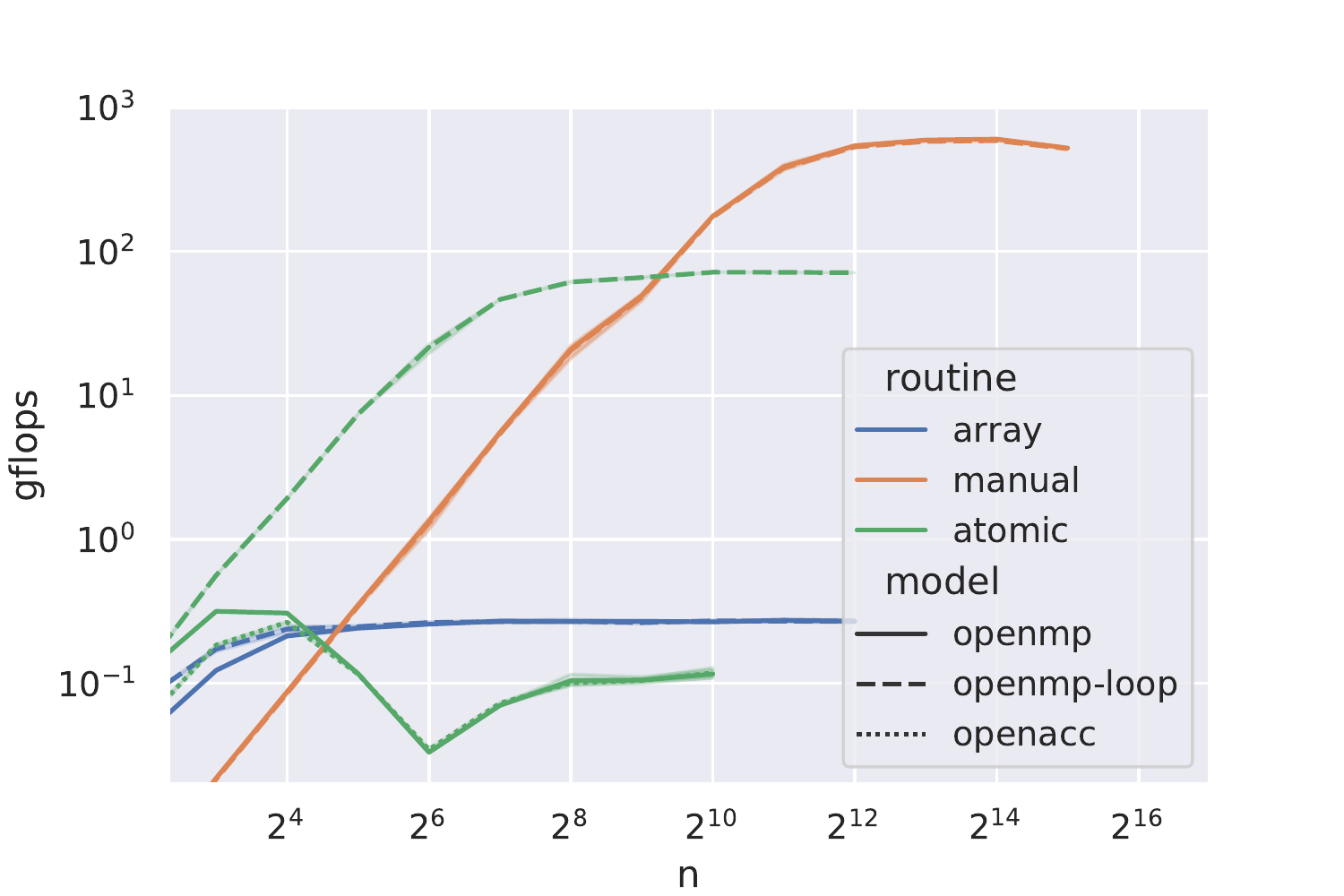}
  \caption{Performance of array reduction with array size of 64 (higher is better).\\
    \textit{Top:} On Skylake CPUs; \textit{Middle:} On A100, where we
    encountered run time errors for $n>2^5$ with OpenMP array reduction and a
    compile error with OpenMP with loops; \textit{Bottom:} On MI100, where for
    $mn^2 \geq 2^{32}$ there appears to be a correctness error due to integer
    overflow on the collapsed loops with the Cray compiler.}
    \label{fig:array-reduction}
\end{figure}

Both the OpenACC 2.7+ and OpenMP 4.5+ specifications support arrays as arguments
to reduction clauses in directives. However, compiler support for these features
varies. With OpenACC, the Cray compiler only provides partial OpenACC 2.6
support and array reductions were not introduced until 2.7. The NVIDIA compiler
also does not support arrays in reduction clauses with OpenACC. With OpenMP both
Cray and NVIDIA compilers support array reductions. However, this feature was
only introduced in NVIDIA's 21.7 release. With Cray the compiler warns, ``An
OpenMP teams construct with an array reduction is limited to a single team.''
With OpenMP, the NVIDIA compiler encounters a run time error as the array size
is increased, or fails to compile when managed memory is used.
Figure~\ref{fig:array-reduction} shows the performance of array reduction on
A100, MI100 and Skylake CPUs respectively where supported. Performance of array
reductions was only competitive with other solutions on Skylake.

\begin{listing}
  \begin{minted}{fortran}
!$acc parallel loop collapse(3)
do i = 1, n
   do j = 1, n
      do k = 1, m
         !$acc atomic
         a(k) = a(k) + x(k,i) * y(k,j)
         !$acc end atomic
      end do
   end do
end do
!$acc end parallel
\end{minted}
  \caption{Array reduction: atomic updates}
  \label{listing:array_reduction_atomic}
\end{listing}

\subsubsection{array reduction with atomics}
With atomic constructs we are able to compile a single version for all
architectures. Compared to the case in Section~\ref{section:fill_array} where
there is no contention, when $n \gg m$, the contention is quite high which
results in a significant performance penalty compared to an optimized reduction
algorithm on the Skylake CPU as seen in the top panel of
Fig.~\ref{fig:array-reduction}. Our results indicate that the implementation
shown in Listing~\ref{listing:array_reduction_atomic} can achieve reasonable
performance on GPUs but not CPUs. We also note that there is an additional
danger with the use of \texttt{collapse} with many loops, the combined iteration
space may manifest a integer overflow for realistic array sizes if 32 bit
integers are used as loop indices even though no individual loop overflows in a
serial implementation.

\subsubsection{generated scalar reductions}
In the case of small arrays another approach is to use a preprocessor that
enables templating and metaprogramming such as \texttt{fypp} to generate
routines that implement a scalar reduction for each element of the array as in
Listings~\ref{listing:fypp_array} and \ref{listing:fypp}. As shown in
Fig.~\ref{fig:array-reduction} this approach achieves good performance on all
architectures. However, this approach works for small arrays, but as array size
is increased runs into several issues either with directive line length (NVIDIA)
or compiler register allocation routines (HPE/Cray). Further it can result in
significant undesirable cognitive and compilation time overhead. In summary no
one method for array reductions is best in all situations, but in this case the
manually generated reductions on scalars with OpenACC give the best overall
cross-platform performance.

\begin{listing}
  \begin{minted}{fortran}
#:def CSV(x,n)
${",".join(f"{x}{i}" for i in range(1, n+1))}$
#:enddef CSV
#:for num_elements in range(2, max_elements+1)
  subroutine reduction${num_elements}$(x, y, a, n, dt)
    integer, parameter :: m = ${num_elements}$
    integer, intent(in) :: n
    real(sp), dimension(m, n), intent(in) :: x, y
    real(sp), intent(out) :: a(m)
    integer :: i,j
    real(dp) :: t0
    real(dp), intent(out) :: dt
#:for i in range(1, num_elements+1)
    real(sp) :: a${i}$
#:endfor
    !$acc data present(x,y)
    t0 = wtime()
#:for i in range(1, num_elements+1)
    a${i}$ = a(${i}$)
#:endfor
    !$acc parallel loop collapse(2) &
    !$acc reduction(+:${CSV("a",num_elements}$)
    do i = 1, n
       do j = 1, n
#:for i in range(1, num_elements+1)
          a${i}$ = a${i}$ + x(${i}$,i) * y(${i}$,j)
#:endfor
       end do
    end do
    !$acc end parallel
#:for i in range(1, num_elements+1)
    a(${i}$) = a${i}$
#:endfor
    dt = wtime() - t0
    !$acc end data
  end subroutine reduction${num_elements}$

#:endfor
\end{minted}
  \caption{Template code for generating reductions with \texttt{fypp}. Listing~\ref{listing:fypp} shows a routine generated for $m=3$.}
  \label{listing:fypp_array}
\end{listing}

\begin{listing}
  \begin{minted}{fortran}
  subroutine reduction3(x, y, a, n, dt)
    integer, parameter :: m = 3
    integer, intent(in) :: n
    real(sp), dimension(m, n), intent(in) :: x, y
    real(sp), intent(out) :: a(m)
    integer :: i,j
    real(dp) :: t0
    real(dp), intent(out) :: dt
    real(sp) :: a1
    real(sp) :: a2
    real(sp) :: a3
    !$acc data present(x,y)
    t0 = wtime()
    a1 = a(1)
    a2 = a(2)
    a3 = a(3)
    !$acc parallel loop collapse(2) &
    !$acc reduction(+:a1,a2,a3)
    do i = 1, n
       do j = 1, n
          a1 = a1 + x(1,i) * y(1,j)
          a2 = a2 + x(2,i) * y(2,j)
          a3 = a3 + x(3,i) * y(3,j)
       end do
    end do
    !$acc end parallel
    a(1) = a1
    a(2) = a2
    a(3) = a3
    dt = wtime() - t0
    !$acc end data
  end subroutine reduction3
\end{minted}
  \caption{Routine for $m=3$ case generated by \texttt{fypp} code in Listing~\ref{listing:fypp_array}.}
  \label{listing:fypp}
\end{listing}

\section{Conclusion and Outlook}

We highlighted several important features of programming for accelerators with
directives that were key for an efficient GPU accelerated port of MFDn. Further
we explored the performance implications of these modifications with CPUs and
with  multiple GPU and compiler vendors.

Avoiding use of private arrays in a production application that has undergone
several years of optimization for multicore CPU platforms was a key challenge.
The conversion of counts to offsets followed by indexing of shared array in a
way that preserves CPU performance while enabling GPU offload was a key pattern
that involved restructuring of many key data structures and routines in the
application.

Our key points for application developers can be summarized as: avoid private
arrays; check compiler diagnostic output to ensure parallel code is in fact
generated; carefully check correctness along with performance; and be mindful of
atomic operations when developing single source code for CPU and GPU
architectures.

Our findings have shown several shortcomings of both the OpenACC and OpenMP
specification/implementation with respect to specialization of code for
different architectures that can hopefully be addressed in future editions of
those specifications and compilers. Finally, we have identified several areas
and motifs that compiler vendors may use to improve their products.

\subsection*{Acknowledgements}
This work is supported by the U.S. Department of Energy (DOE) under Award
Nos.~DE-FG02-95ER40934 and DE-SC0018223 (SciDAC/NUCLEI), and by the DOE Office
of Science, Office of Workforce Development for Teachers and Scientists, Office
of Science Graduate Student Research (SCGSR) program (administered by the Oak
Ridge Institute for Science and Education (ORISE), managed by ORAU under
contract number DE‐SC0014664). This research used resources of the National
Energy Research Scientific Computing Center (NERSC), a DOE Office of Science
User Facility located at Lawrence Berkeley National Laboratory, operated under
Contract No. DE-AC02-05CH11231, as well as resources of the Oak Ridge Leadership
Computing Facility at the Oak Ridge National Laboratory, which is supported by
the DOE Office of Science under Contract No. DE-AC05-00OR22725.

\bibliographystyle{splncs_pinilla}
\bibliography{refs}
\end{document}